%% file: main.tex
\documentclass[12pt]{article}

\input{setup/packages.tex}

\begin{document}

\thispagestyle{empty}
\setlength\headheight{0pt} 

\begin{center}

\begin{center}
\includegraphics[width=0.50\linewidth]{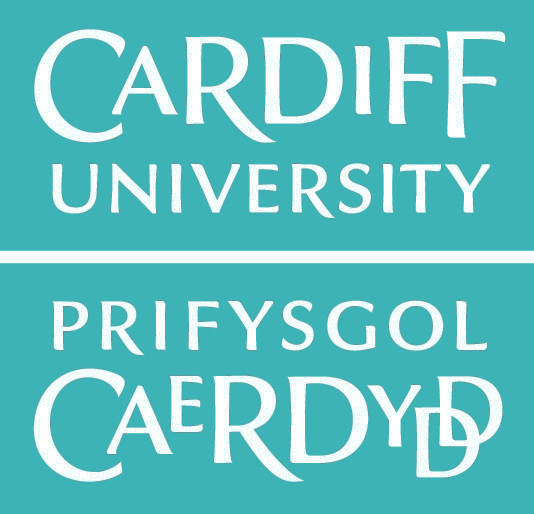}            
\end{center}	

        \vspace{0.25cm}
        {\scshape\Large Cardiff University\par}
        \vspace{0.25cm}
        {\scshape\Large School of Computer Science and Informatics\par}
        \vspace{0.5cm}
        -----------------------------------------------------------\\
        \vspace{0.5cm}
        {\LARGE\bfseries Real-time error correction and performance aid for MIDI instruments\par}
        
        \vspace{0.5cm}
        {\Large BSc Computer Science Dissertation\par}
        \large
        6 June 2020

\vspace{2cm}
\begin{multicols}{3}
Author:\par
Georgi Marinov \\
Supervisor:\par
Dr. Kirill Sidorov \\
Moderator:\par
Prof. Dave Marshall \\
\end{multicols}
\end{center}

\clearpage
\restoregeometry
\justify

\section*{Acknowledgements}
I would like to thank my supervisor, Dr. Kirill Sidorov, for his invaluable ideas and observations, which proved crucial for the project.\par
I would like to acknowledge my moderator, Professor Dave Marshall, for inspiring my interest in DSP and combining computer science with music, as well as his digital music representation knowledge and highly structured resources.\par
I'd also like to extend my gratitude to Sander Dieleman from DeepMind, for his inspiring artificial intelligence research, especially in music generation.\par
Special thanks for the insightful and practical suggestions of Ian Simon and Monica Dinculescu of the Magenta research team from Google Brain, further steering me towards exploring music generation for error correction and helping me bring the results of the research as close to the user as possible - in the browser.\par
The completion of my dissertation would not have been possible without the support and nurturing of my family and the calming presence of my dog.\par
Special thanks to Kalina Anastasova for her profound belief in my abilities and for helping me maintain my sanity.\par
Lastly, thanks should also go to Dr. Károly Zsolnai-Fehér, creator of the inspiring YouTube channel "2 Minute Papers", for promoting the amazing Weights and Biases platform, which made training and visualising countless model prototypes throughout the project possible.\par

\pagebreak

\listoffigures
\listoftables
\pagebreak

\tableofcontents
\pagebreak
\section*{Abstract}
Making a slight mistake during live music performance can easily be spotted by an astute listener, even if the performance is an improvisation or an unfamiliar piece. An example might be a highly dissonant chord played by mistake in a classical-era sonata, or a sudden off-key note in a recurring motif. The problem of identifying and correcting such errors can be approached with artificial intelligence - if a trained human can easily do it, maybe a computer can be trained to spot the errors quickly and just as accurately. \par

The ability to identify and auto-correct errors in real-time would be not only extremely useful to performing musicians, but also a valuable asset for producers, allowing much fewer overdubs and re-recording of takes due to small imperfections.\par

This paper examines state-of-the-art solutions to related problems and explores novel solutions for music error detection and correction, focusing on their real-time applicability.
The explored approaches consider error detection through music context and theory, as well as supervised learning models with no predefined musical information or rules, trained on appropriate datasets. \par

Focusing purely on correcting musical errors, the presented solutions operate on a high-level representation of the audio (MIDI) instead of the raw audio domain, taking input from an electronic instrument (MIDI keyboard/piano) and altering it when needed before it is sent to the sampler. \par

This work proposes multiple general recurrent neural network designs for real-time error correction and performance aid for MIDI instruments, discusses the results, limitations, and possible future improvements. It also emphasizes on making the research results easily accessible to the end user - music enthusiasts, producers and performers - by using the latest artificial intelligence platforms and tools.\par

\pagebreak

\section{Introduction}
Solving the problem of instrument error correction in a live musical performance would usually entail altering the audio produced by the instrument in real-time and outputting the corrected audio back. This end-to-end pipeline introduces other complicated problems such as audio signal separation and automatic music transcription. There are promising recent machine learning solutions for those problems by \cite{hawthorne2017onsets}, however this project will solely focus on error detection and correction in real-time, which is unexplored so far.\par 

For this purpose, the problem will be reduced to error correction for MIDI instruments, as the MIDI musical representation enables working directly with digitized musical information such as note pitches and velocities. It also allows easy integration of the proposed error correction software in any musical system - musicians will be able to test the results of this research by using a simple MIDI plugin.

\subsection{Approaches}
There are multiple ways to identify an error or a mistake in a performance.
One is to derive the musical context of the performance and check if a note belongs to it - notes that do not belong in the key and scale of the performance are generally sounding 'off'.
Other methods may include determining the harmony of the note with the preceding ones, for which again musical context and music theory can be used.
There are also artificial intelligence methods, which learn the relationships between the notes without being provided with any music theory, by training on large amounts of data. The proposed AI method can be used to judge how well a note fits with previously played notes, thus identifying potential mistakes.\par

For correcting the identified mistake we have a similar choice of methods. Using derived musical context, the off-sounding note can be replaced by an adjacent note which belongs to the determined key and scale. This would again require rules, based on music theory. If, instead, AI is used to estimate how fitting the current note is in relation to previous notes, it can also be used to find the best-fitting note close to the played one to replace it with, without predefined musical rules.\par

\pagebreak

\subsubsection{Error detection}
The task of detecting an error in a musical performance is a classical problem that can be approached with machine learning. Usually, a person can easily identify a 'wrong' note in a performance, along with its perceived magnitude or likelihood of being a mistake, but often the exact reason for such a perception is not apparent. Trained and experienced musicians can even better recognize a supposed mistake and sometimes explain the mistake through music theory.\par

There are also cases where even musicians cannot unambiguously determine if a note is a mistake in real time. Playing a recurring motif differently, or passing through an off-key note may be intentional, depending on the characteristic of the performance and the style of the artist. Errors in jazz, progressive or non-diatonic music, for instance, can be more difficult to spot compared to genres in which repetition of patterns and more strict adherence to chords and notes in a certain key or scale are common, such as classical music.\par

To understand why a note is 'wrong', musicians could use music theory concepts to analyze the musical context, but they need to listen the performance to its end, or listen to it multiple times in order to be able to do so. Even with very complicated music, professionals can give a set of reasons as to why a particular note does not fit, or how it could sound better. In the most extreme cases of contemporary jazz music, some musicians, such as the popular jazz enthusiast and YouTuber Adam Neely, are able to analyze a piece of music, re-harmonize it, or make suggestions on how to give it a particular type of sound. Deriving a musical context within a more complex piece of music is possible, but it might be very challenging and time consuming. Without the ability to look ahead, however, it might even not be possible, as music performances have a structure and multiple different contexts.\par

This ability to recognize mistakes and suggest improvements is made possible not only by following music theory rules or context, but mostly by following flair that has been acquired through years of experience, learning and listening to music. This is what gives humans their intuition in detecting "mistakes" in music, and this is what artificial intelligence might be able to recreate. This acquired intuition will allow digital detection of errors in real-time, similar to the human perception.\par

\subsubsection{Error correction}
Assume that despite the challenges of error detection we have estimated it correctly. An error correction model aims to find the note that was originally intended to be played, and reproduces it. There are different approaches to correct an error. One can attempt to fit the wrong note in the derived context of the musical piece. Another strategy is to explore adjacent notes, see whether they are a better fit than the initially played wrong note, and check the likelihood that some of them matches the intention of the player. Of course, the choice of notes is constrained by the physical limits of a player's reach on a keyboard. An error is most often adjacent key, but could also allow a range of up to 3 semitones for this check.\par

One can never be sure of what the intention of the player was, or whether he truly made a mistake, but acting on it in a reasonable confidence interval might catch most true errors while avoiding the correction of intentional off-key notes.
\subsubsection{Performance aid}
The final approach is to use AI model for music generation to solve the error-correction problem.
Training a general AI, which attempts to continue the performance based on previous notes can solve both error detection and error correction problems by comparing the current playing notes with the prediction.
If the note that the user plays is very unlikely to be played by the AI, it can be considered a mistake. The correct note can then simply be what the AI would most likely play, that is again limited to notes that are physically close on the piano.

This approach also allows something more than error correction - performance aid. Instead of altering the notes only when the AI decides that the player has made an error, the model could occasionally just play better fitting notes than the currently played ones. This would broaden the use cases of such software, as composers could also use this, for exploring new ideas influenced by the AI. 

The performance aid also allows configuring how often should the AI override the user.
A novice piano player might use high level of performance aid from the AI model when playing, while an experienced piano player might benefit from only a slight aid, for example when he wants to play less carefully and still sound "good".

\section{Literature Review}
While there has been good amount of research on related problems such as musical key and scale estimation (music context) by \cite{contextestimating1}, \cite{contextestimating2} and \cite{context3}, it seems like real-time music error correction is yet a niche research topic. 
My initial observations about error correction through musical context derivation have been altered after learning more on the subject, namely because of the requirements and performance of such estimators. They often need the whole musical piece and are not feasible to be applied in real-time error correction software with such a strong run time constraint - a slight delay of couple of milliseconds will be noticeable in a live performance. \par
The proposed AI models can be interpolated during inference calculations, which is often applied in other audio applications, as done by \cite{audiodomain}, or used in combination with partitioned convolution and windowed Fast Fourier Transform implementation to achieve near zero latency with the model performance as done by \cite{wavenetaudiodomain}, but attempting to implement or improve on state-of-the-art musical context estimators is not within the scope of this project.
Furthermore, those context estimators are for more complicated music recordings, operating in the raw audio domain, which is also not in scope. \par

On the contrary, recent research in artificial music generation field has provided techniques that can also be applied in attempt to solve the error correction problem.
Until recent years the subject of music generation was explored mostly through composition and producing scores, without taking human expressiveness and performances into account, but recently there's been improvements on expressive music generation by \cite{performancernn3} and similar research directly in the audio domain, for example MelGAN by \cite{melganWaveformSynthesis} and \cite{audiodomain}.

The idea to continue a musical sequence has been around for a while and there have been some really astonishing results achieved by Google Brain's Magenta team.
PerformanceRNN from \cite{performancernn2} and Transformer by \cite{transformer} are state-of-the-art music generation tools that can produce virtuoso performances and music with long term structure.\par

PerformanceRNN is a relatively simple RNN that uses a LSTM cell, but the end result is simply inspiring. 
This research provided valuable insights on data representation and music generation model architecture, which were adopted to some degree in this paper for correcting mistakes in a real performance.
Please check PerformanceRNN and their interactive web demo!\footnote{\url{https://magenta.tensorflow.org/performance-rnn}}

Transformer seems very fitting to the problem description, because it is based on attention which outperforms LSTMs - "Attention is all you need" by \cite{attentiongoogle}, however the attention based model is very slow and memory intensive. Even with the state-of-the-art models and optimisations used in related research by Deepmind and Google Brain, attention based models are not yet feasible for real-time applications. The performance of the model proposed in this paper could most likely be improved when applying attention for this problem is possible.

Another paper of significance for this research is PianoGenie by \cite{pianogenie}, which uses Encoder/Decoder model to transform a sequence of 8 buttons into a piano performance, and is capable of running in real-time with no noticeable latency. Unfortunately it is not applicable for error-correction, as it does not allow for fine tonality control over the generated performances, having only 8 input classes.
The sequence-to-sequence model could potentially be used for achieving error-correction or performance aid, but with higher number of input types (potentially the full 88 piano keys) the model will quickly become too compute-intensive for real-time purposes.
\pagebreak

\section{Method}
This section will go over the discussed approaches for error detection and error correction in detail, focusing more on the artificial intelligence solutions.
\subsection{Error correction by musical context}
Starting with the basic problem of error correction when we have the musical context - the detection can be done by checking whether the current note fits the current context.
The method for doing this is simple:
\begin{enumerate}
	\item Calculate the note's pitch relative to the key (from the context). The key is represented as an integer from 0 to 11 inclusive, where 0 is C, 1 is C\# and so on. The note's is represented in the MIDI standard as an integer from 0 to 127, where 60 is C4 (middle C).
	\item Depending on the key and mode\footnote{\url{https://en.wikipedia.org/wiki/Mode_(music)}} the note can be considered off-key or a mistake if it does not belong to the 7 notes of the target scale.
	\item The note can be replaced by the closest note within the scale.
\end{enumerate}
The last step is the trivial error correction method which is not suitable for complicated pieces or e.g. jazz music, but working very well on less complicated music.
The developed VST plugin incorporates this simple approach as a base solution, achieving real-time error correction by musical context. 
 \subsubsection{Automatic context detection}
 This method could be further improved by applying automatic context detection, for which there are multiple existing solutions examined. Improvement was not pursued, as those solutions are crafted to operate mostly on the raw audio domain and are generally more complex than needed for the project, as they also consider multiple instruments. Another problem is that the estimated key and scale could be suddenly changed, as it often happens - the user will have to "stop" the error correction when transitioning to avoid staying in the same key for example.
\subsection{Error detection with Machine Learning}
 Artificial intelligence can be used with the sole purpose of identifying errors in a live music performance. Since every note can be either a mistake or not, we can formulate the problem as a statistical binary classification.
 Binary classification is a well studied supervised learning problem in machine learning, used to categorize new probabilistic observations in two categories, in this case a correct note or a mistake.
 To do binary classification (or any supervised learning task), a dataset of many example inputs with their categorized outputs is needed. For this task, we will need a dataset consisting of MIDI sequences and labeled "mistakes" in them.
\subsubsection{Dataset}
A dataset of identified mistakes within MIDI sequences does not exist, and probably will not be gathered - for such dataset to exist, people would have manually identified errors in millions of MIDI recordings and collected the labeled data.\par
Fortunately, datasets of many midi recordings do exist, like the Lakh MIDI Dataset (\cite{lakhdata}) or the fitting Maestro dataset (\cite{hawthorne2018enabling}).
The Lakh MIDI dataset is partially matched to songs from the Million Songs Dataset \footnote{\url{http://millionsongdataset.com/}} and consists of produced MIDI files with multiple instruments and unlabeled performances. The Maestro dataset was collected from the international piano e-competition and is of very high quality - "During each installment of the competition virtuoso pianists perform on Yamaha Disklaviers which, in addition to being concert-quality acoustic grand pianos, utilize an integrated high-precision MIDI capture and playback system. Recorded MIDI data is of sufficient fidelity to allow the audition stage of the competition to be judged remotely by listening to contestant performances reproduced over the wire on another Disklavier instrument."\footnote{\url{https://magenta.tensorflow.org/datasets/maestro}}
There are no errors in those MIDI datasets, or at least they are not recorded or produced to have such. Participants in the piano competition are all professional performers and is highly unlikely to identify mistakes in their performances.\par
  
  \subsubsection{Dataset generation}
  Since the dataset contains no errors, they could be artificially generated.\par
  Suppose we have a section of a performance where only notes within a certain keys are played. If we change one of the notes in the middle part of the sequence to be off key it will look and also sound like an error. Also a pattern repeating multiple times can be "ruined" by altering some of the notes in it.
  
  Unfortunately, those observations are not necessarily correct - While repetition is a good measure, as "repetition is legitimizing"\footnote{Quote by YouTuber and musician Adam Neely}, one of the key principles in progressive music improvisation is to alter the repeated sequences increasingly. When trying to generate errors in some of the performance recordings from the international piano e-competition, I have failed to recognize the vast majority of the "mistakes", as they sounded equally "right" and interesting. There's a recognizable difference for most of the generated examples, but which notes are mistakes seems unclear. But, "talking about music is like dancing about architecture"\footnote{Quote by Frank Zappa, among others.}, you can check some of the results of the artificial error generation yourselves.\footnote{\url{https://midi-tune-test.shefa.xyz/} - you will hear two versions of several performances - can you tell the originals from the ones with the errors?}
  
  Even if those assumptions were true, it is irrelevant - every artificially generated mistake, no matter how complex the model for generating them is, will introduce extreme bias in the dataset, as mistakes in real performances have variety of properties that cannot be modeled by simple logic - if they could, this logic could be applied in solving the problem instead of building the dataset. \par
  Musicians also seem to have hard time making mistakes on purpose - as a musician myself, it is very awkward to recreate a believable mistake intentionally.\par
  In conclusion, generating errors artificially to train a model to detect those errors is a hard problem in itself, and will likely cause poor model performance. There are better uses and representation of the mentioned datasets in the next section.

\subsection{Error correction and performance aid}
 Another approach to error correction by utilizing supervised learning is to use a Neural Network to predict the continuation of the note sequence so far and compare the results with the notes from the user input. 
 While this might seem unusual at first, there are some key observations that make the idea more appropriate than the binary classification:
\begin{enumerate}
    \item There's no need for generating errors in the dataset artificially $\rightarrow$ no bias.
    \item No explicit music theory used, the model will attempt to learn the relationships of notes and sequences from the dataset.
    \item Unified error detection and error correction solution - the model predictions for continuing the input sequence will allow to both check if the current note can be classified as a mistake (if its probability of being in the continuation is too low) and which of the adjacent notes is fitting the input sequence better (most likely the intended note).
\end{enumerate}
\subsubsection{Architecture}
As briefly explained in the literature review, the idea for solving error correction with music generation comes heavily from the PerformanceRNN and the used model there.
The proposed model is a simple Recurrent Neural Network using long short-term memory cell (LSTM), which will take a sequence of notes as input and output probabilities for each of the 88 piano pitches being the next one in the sequence.
Considerations and attempts have been made at using attention based RNN, similar to Transformer, but the simple RNN produced great results.
\subsubsection{Dataset}
\label{sec:data}
The used dataset is the aforementioned Maestro dataset gathered from Yamaha's international piano e-competition. The MIDI files contain a single instrument - the piano - and are in standard MIDI message form.\par

\begin{figure}[ht!]
 	\centering
 	\caption{Velocity values distribution}
 	\includegraphics[width=0.5\linewidth]{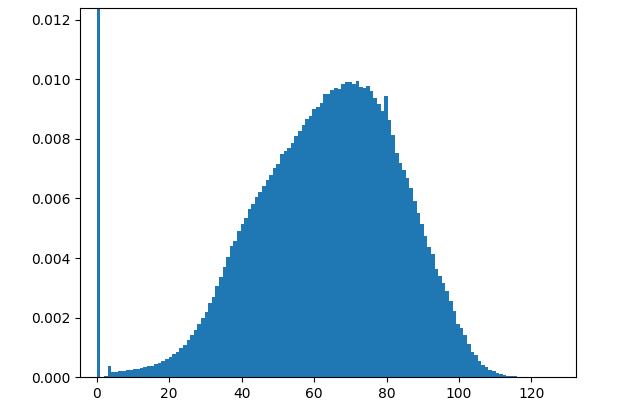}
 	\label{lab:velocities_distrib}
 \end{figure}

\begin{figure}[h!]
 	\centering
 	\caption{Delta values distribution}
 	\includegraphics[width=0.5\linewidth]{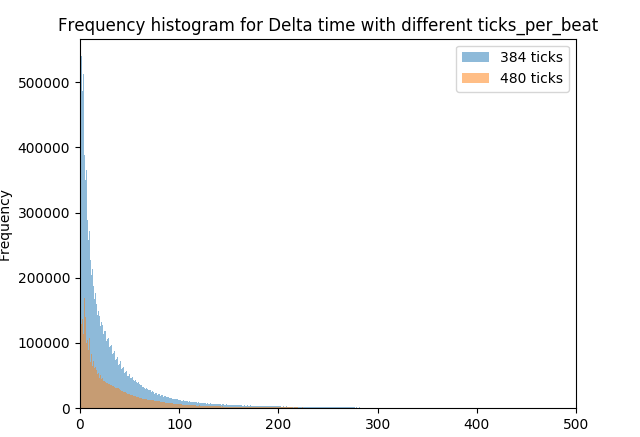}
 	\label{lab:delta_stuff}
 \end{figure}

The original dataset has been used to create multiple, increasingly complex datasets to train and test models with different number of parameters:
\begin{enumerate}
    \item Basic dataset - comprised of only the pitches of the note\_on MIDI events - a sequence of pitches with no additional information.
    \item Velocity dataset - in addition to pitches, note velocities are also stored in this representation. To reduce model complexity, velocity values have been categorized in 8 buckets, as show in in table \ref{tab:vvalues}. Note velocities in the dataset are normally distributed, as shown in figure \ref{lab:velocities_distrib}. The values for the bins have been picked in such a way to preserve this distribution - most of the notes are in the medium velocity category bin, while only a small subset is in the loudest category. The category bins are also related to the notion of dynamics in music\footnote{\url{https://en.wikipedia.org/wiki/Dynamics_(music)}}, and are similar to values used in music notation programs.

\begin{table}[ht!]
\centering
    
	\caption{Velocity buckets}
	\begin{tabular}{ |l|c|c|}	
		\hline		
		\textbf{Music dynamic notation} & \textbf{Velocity} & \textbf{Bucket} \\ \hline
		fortississimo   fff very very loud & 86 $\leq$ V <128 & 7  \\ \hline
        fortissimo	ff	very loud & 78 $\leq$ V <86 & 6  \\ \hline
        forte	f	loud & 71 $\leq$ V <78 & 5  \\ \hline
        mezzo-forte	mf	average & 65 $\leq$ V <71 & 4  \\ \hline
        mezzo-piano	mp & 58 $\leq$ V <65 & 3    \\ \hline
        piano	p	soft & 50 $\leq$ V <58 & 2  \\ \hline 
        pianissimo	pp	very soft & 40 $\leq$ V <50 & 1 \\ \hline
        pianississimo	ppp	very very soft & 0 $\leq$ V <40 & 0  \\ \hline
	\end{tabular}
	\label{tab:vvalues}
\end{table}

 \item Basic delta dataset - further extending the complexity of the dataset, the time between the note\_on events is added. In order to reduce complexity and normalize this time delta data, it is categorized and placed in 12 bins, as shown in table \ref{tab:dvalues}. The scale is logarithmic and is again matching the data distribution and human perception. Some MIDI files have different measure of time indicated by the 'ticks\_per\_beat' property as shown in figure \ref{lab:delta_stuff}, which has been taken into account by scaling the deltas accordingly.
 \item Full dataset - the final level of complexity is by using the MIDI data as it is provided. This means having separate events for notes being played and being stopped (note\_on and note\_off events). The pitch information along with the velocity and delta time categorized attributes are kept.
\end{enumerate}

 \begin{table}[h!]
\centering
	\caption{Delta time buckets}
	\begin{tabular}{ |l|c|c|}	
		\hline		
		\textbf{Delta time} & \textbf{Bucket} \\ \hline
		1024 $\leq$ D & 11  \\ \hline
		512 $\leq$ D  <1024 & 10  \\ \hline
		256 $\leq$ D  <512 & 9  \\ \hline
		128 $\leq$ D  <256 & 8  \\ \hline
        64 $\leq$ D <128 & 7  \\ \hline
        32 $\leq$ D <64 & 6  \\ \hline
        16 $\leq$ D <32 & 5  \\ \hline
        8 $\leq$ D <16 & 4    \\ \hline
        4 $\leq$ D <8 & 3  \\ \hline 
        2 $\leq$ D <4 & 2   \\ \hline
        1 $\leq$ D <2 & 1  \\ \hline
        D = 0 & 0  \\ \hline
	\end{tabular}
	\label{tab:dvalues}
\end{table} 

\subsubsection{Data encoding}
\label{sec:encoding}
The input of the neural network is a sequence of notes represented in some of the formats discussed above. Since we're dealing with different types of data, before feeding them to the RNN, the sequences must be encoded.
When dealing with categorical data, in general one-hot encoding is preferred to ordinal encoding. However, one-hot encoding requires enormous amount of memory to represent all sequences, that's why models with different encoding were tested, including:
\begin{enumerate}
    \item One-hot encoded both input and output categorical variables - only feasible for small training size or small number of epochs, without breaking the bank.
    \item Ordinal input, One-hot encoded output - saves enough memory to be able to use whole dataset for training, but model converges very slowly.
    \item Ordinal encoding for both input and output by using the sparse categorical cross-entropy loss function - allows to train models with much fewer RAM consumption, with similar performance to one-hot encoding.
\end{enumerate}
\subsubsection{Model}
\begin{figure}[ht]
 	\centering
 	\caption{RNN model}
 	\includegraphics[scale=0.4]{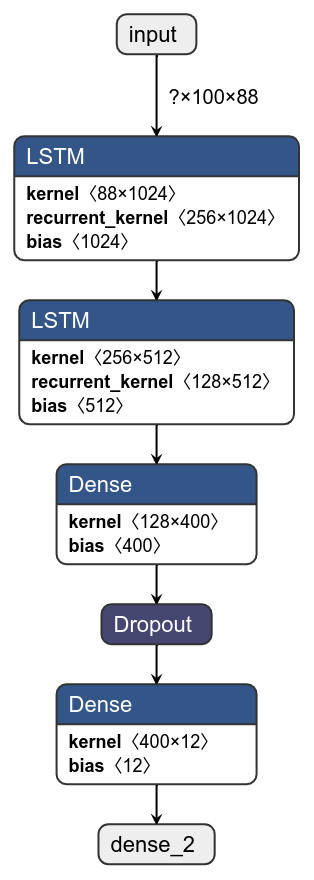}
 	\label{lab:model}
 \end{figure}

The final model is a very simple RNN consisting of 2 LSTM cells followed by a dense hidden layer and an output layer, as shown in figure \ref{lab:model}.
\subsubsection{Training}
Training the model is a relatively easy process but tuning the hyper-parameters can be very difficult and consume a lot of resources.

The platform Weights and Biases was used along with powerful GPU computing agents from Google Cloud Platform to conduct many training runs of the model with different combination of hyperparameters, such as:
\begin{itemize}
    \item Input sequence length - the number of notes the model will "look back"
    \item Model dataset and encoding types - from \hyperref[sec:data]{section 3.3.2} and \hyperref[sec:encoding]{section 3.3.3}
    \item Layer sizes, Epochs, Batch size, Dropout, Learn rate - standard deep learning hyperparameters
\end{itemize}

\subsection{User interaction}
In order to make the results of this research accessible to musicians without any special equipment or software, the interaction with the model can be done via:
\begin{itemize}
    \item a simple VST plugin which can be loaded in any Digital Audio Workstation
    \item a standalone application for Linux and Windows
    \item an interactive web demo, running on any device - ideally, anyone with a MIDI keyboard (or digital piano) should be able to plug it in their PC or even mobile device, open their browser and play music, interacting with the trained AI model.
\end{itemize}

\pagebreak

\section{Results}
\subsection{VST plugin}
The result of the first approach to the error correction is a very simple VST plugin that can be used in any Digital Audio Workstation.
The plugin can be placed in the plugin chain before the sampler, resulting in the desired behaviour of real-time error correction by adjusting the notes that are not within the current musical context.
The context is defined as a combination of key (root note) and mode, as detailed in the methods section.

The current context is changed by the user, by providing the information on the simple plugin UI.
Despite the simplicity of the model, the end result is quite satisfying and easy enough to use.
I was able to "improvise" on some backing tracks without having to learn and adhere to the key.

\begin{figure}[h]
\caption{Midi-tune VST in Reaper}
\includegraphics[width=0.99\linewidth]{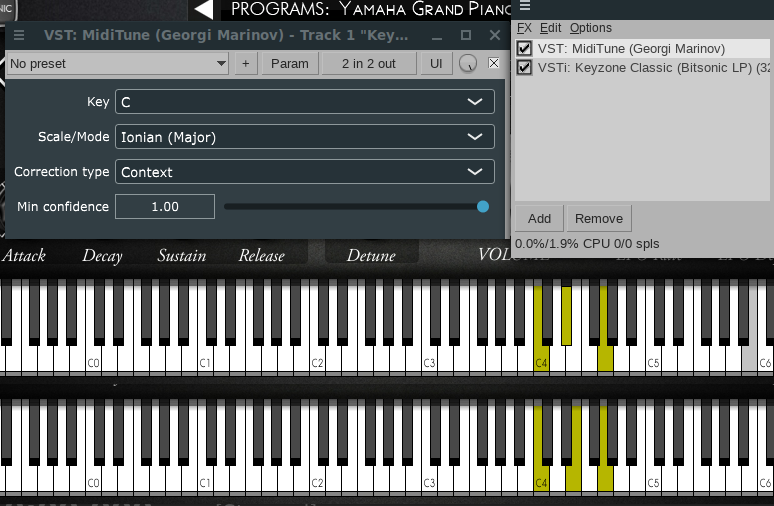}
\end{figure}
Shown on the figure is a C minor chord corrected in the key of C major instantly.
The plugin is developed in \verb!C++! with the JUCE framework and has universal VST plugin integrating in all DAWs and pre-built binaries for Windows and Linux.
The Linux version also has integrated the RNN model (the correction type dropdown).
\subsection{Recurrent Neural Network models}
\subsubsection{Simple data representation model}

This is the report on the results after running hyperparameter sweep on the basic model with simplified input (only pitches).
After 200 runs on GCP machines for a period of 3 days, the accuracy was around 0.84 and the correlation between hyper-parameters and model performance has been sufficiently analysed.
You can find more information on \href{https://app.wandb.ai/shefa/midi-tune-ML/reports/Basic-Model-Hyperparameters-200-Runs--VmlldzoxMzExODE}{the report page in weights and biases (click here).}
\begin{figure}[h]
\caption{Basic model hyperparameter optimization 200 runs}
\includegraphics[width=0.49\linewidth]{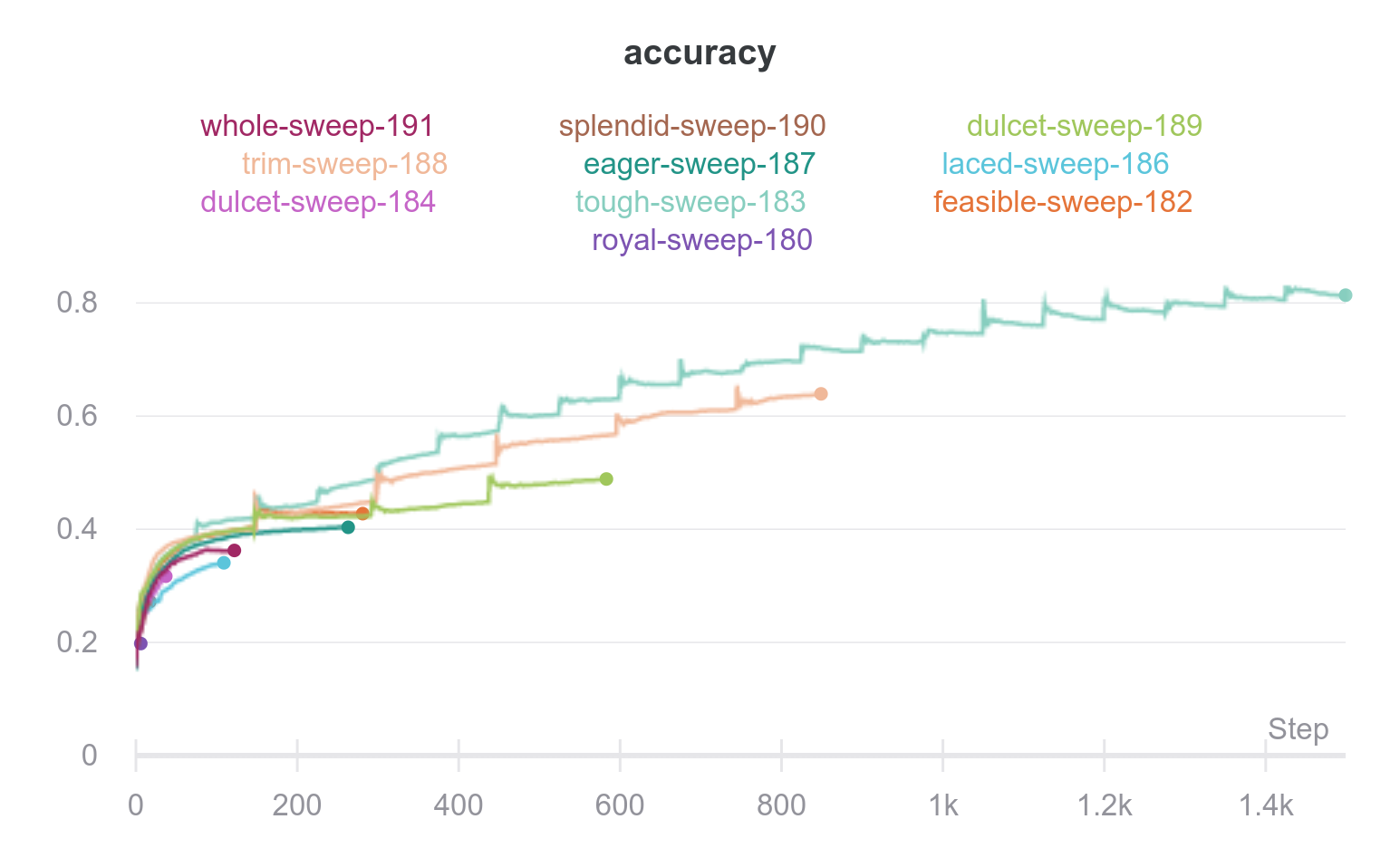}
\includegraphics[width=0.49\linewidth]{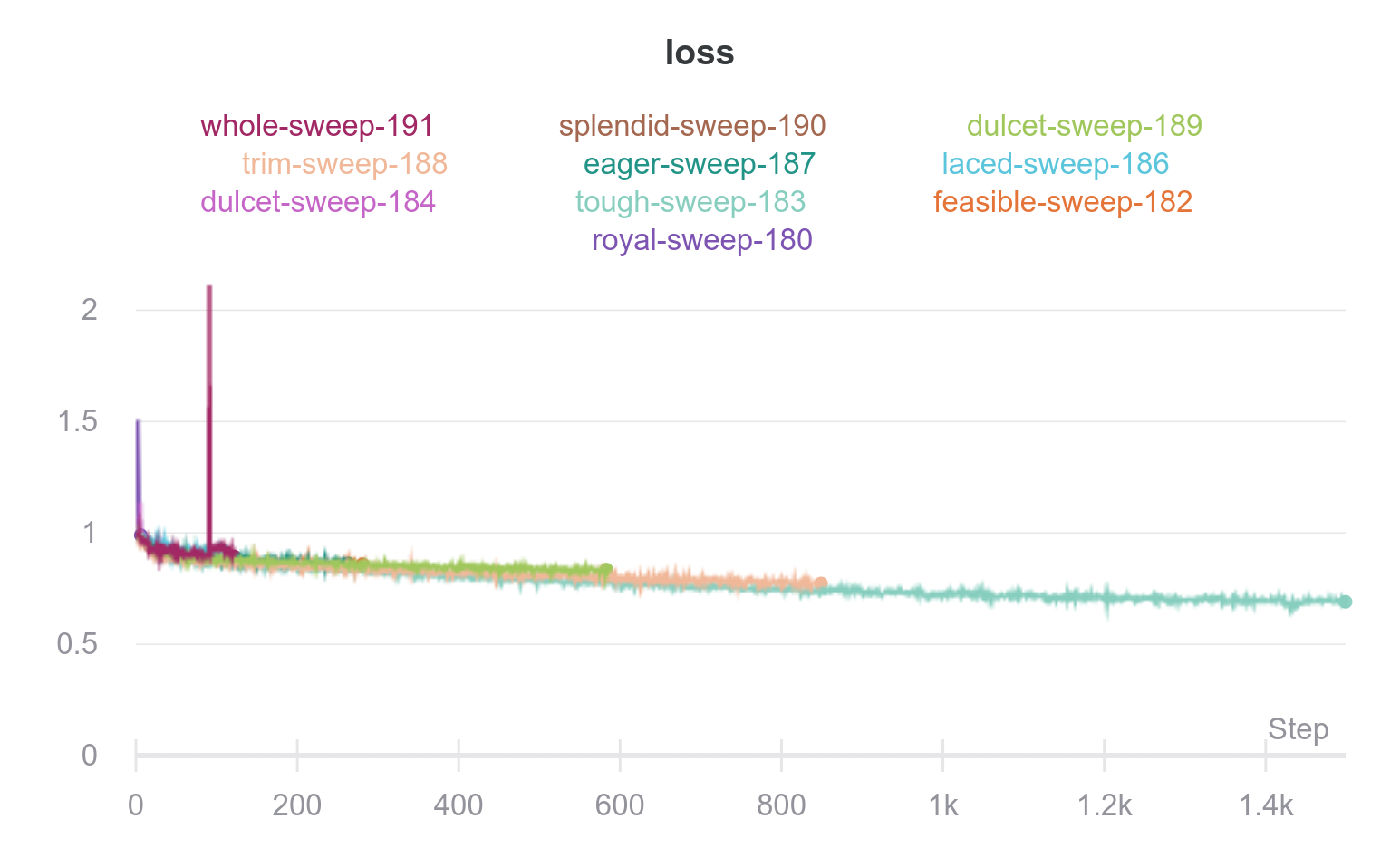}
\includegraphics[width=1.0\linewidth]{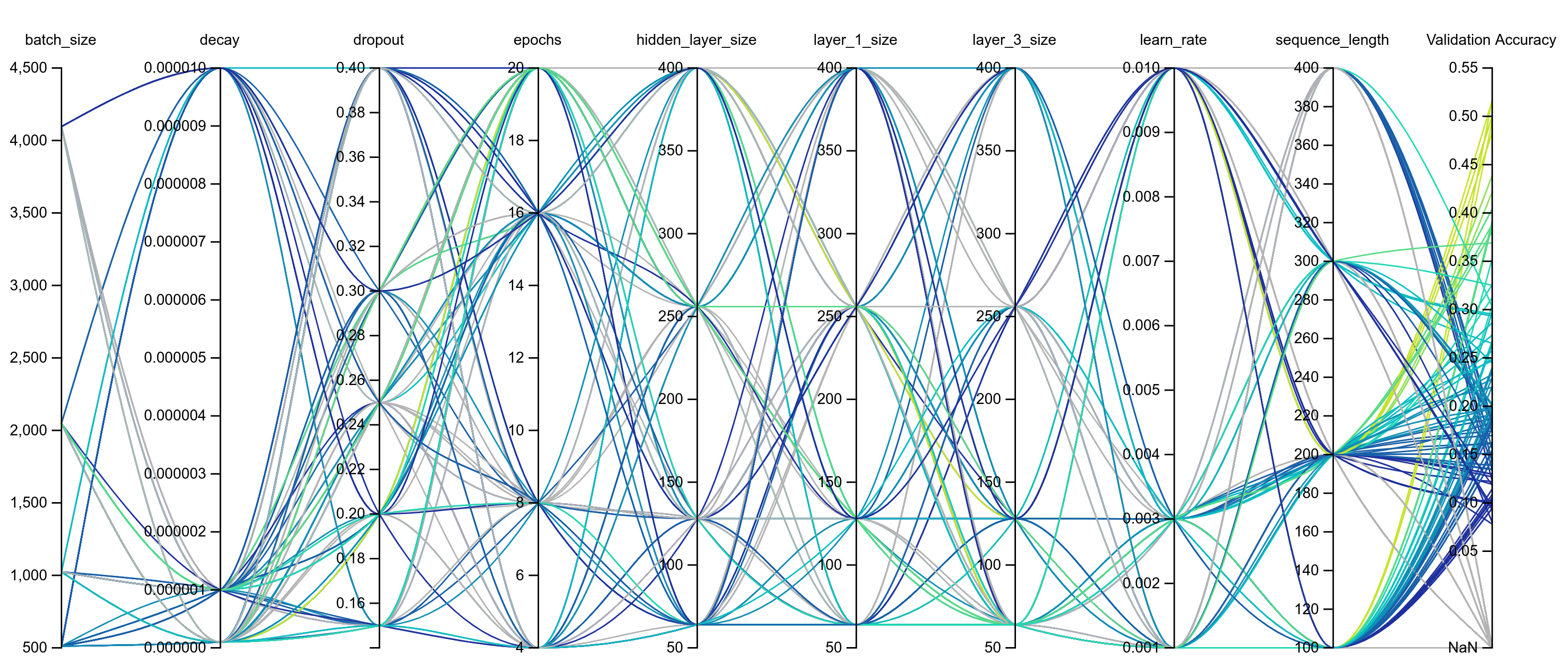}
\end{figure}

\subsubsection{Complex (full) data representation model}
This is the report from running hyerparameter sweep on the full model with the most complex input (pitches, velocities and deltas).
Unlike the simple model, this model took way shorter time to converge and get >80\% accuracy.
Similar to the other model, you can find more information on the correlation between hyperparameter values and model performance and other interesting interactive information in \href{https://app.wandb.ai/shefa/midi-tune-ML/reports/Complex-Model-Hyperparameters-30-runs--VmlldzoxMzI4MzM}{the report page in weights and biases (click here).}
\begin{figure}[h]
\caption{Complex model hyperparameter optimization 30 runs}
\includegraphics[width=0.49\linewidth]{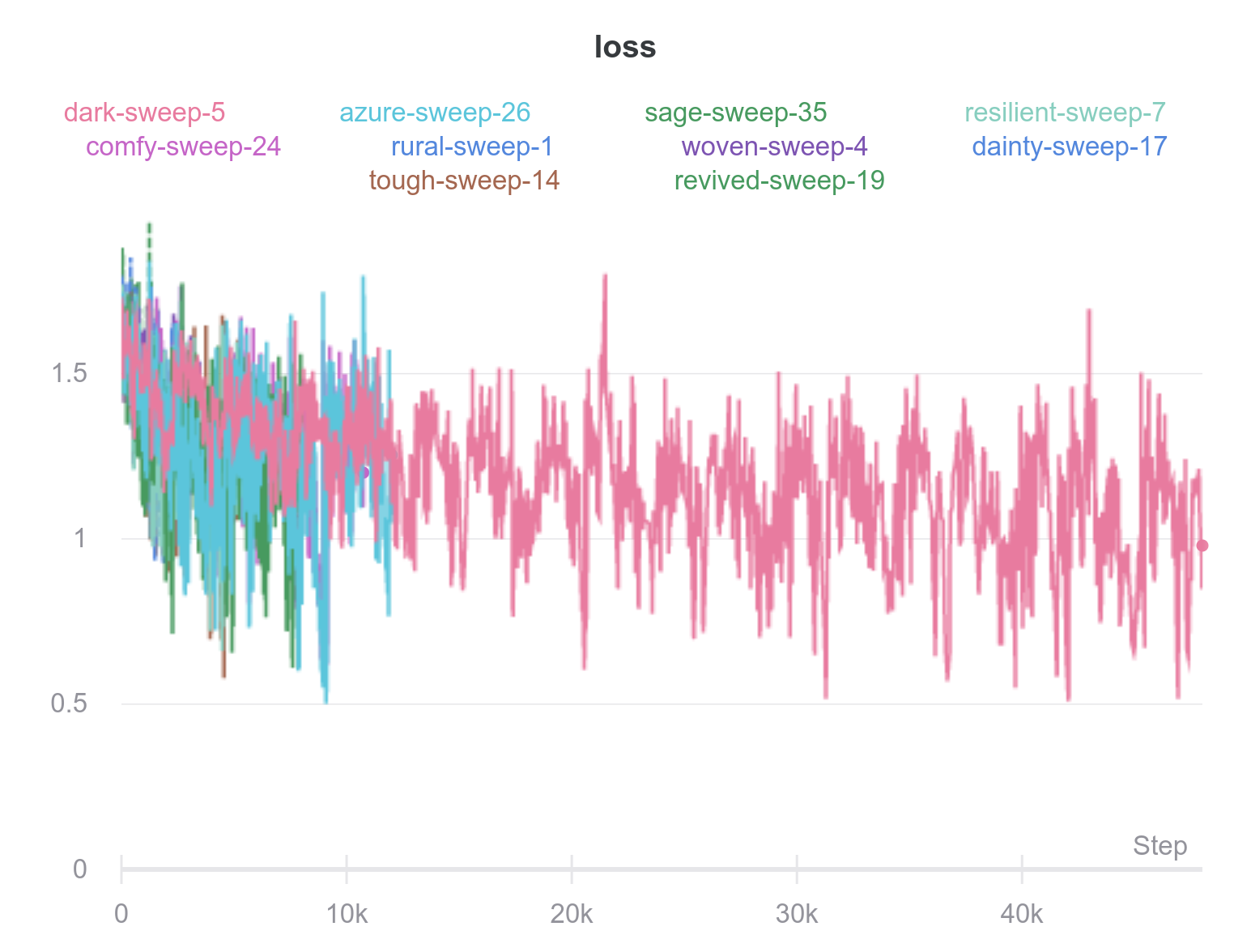}
\includegraphics[width=0.49\linewidth]{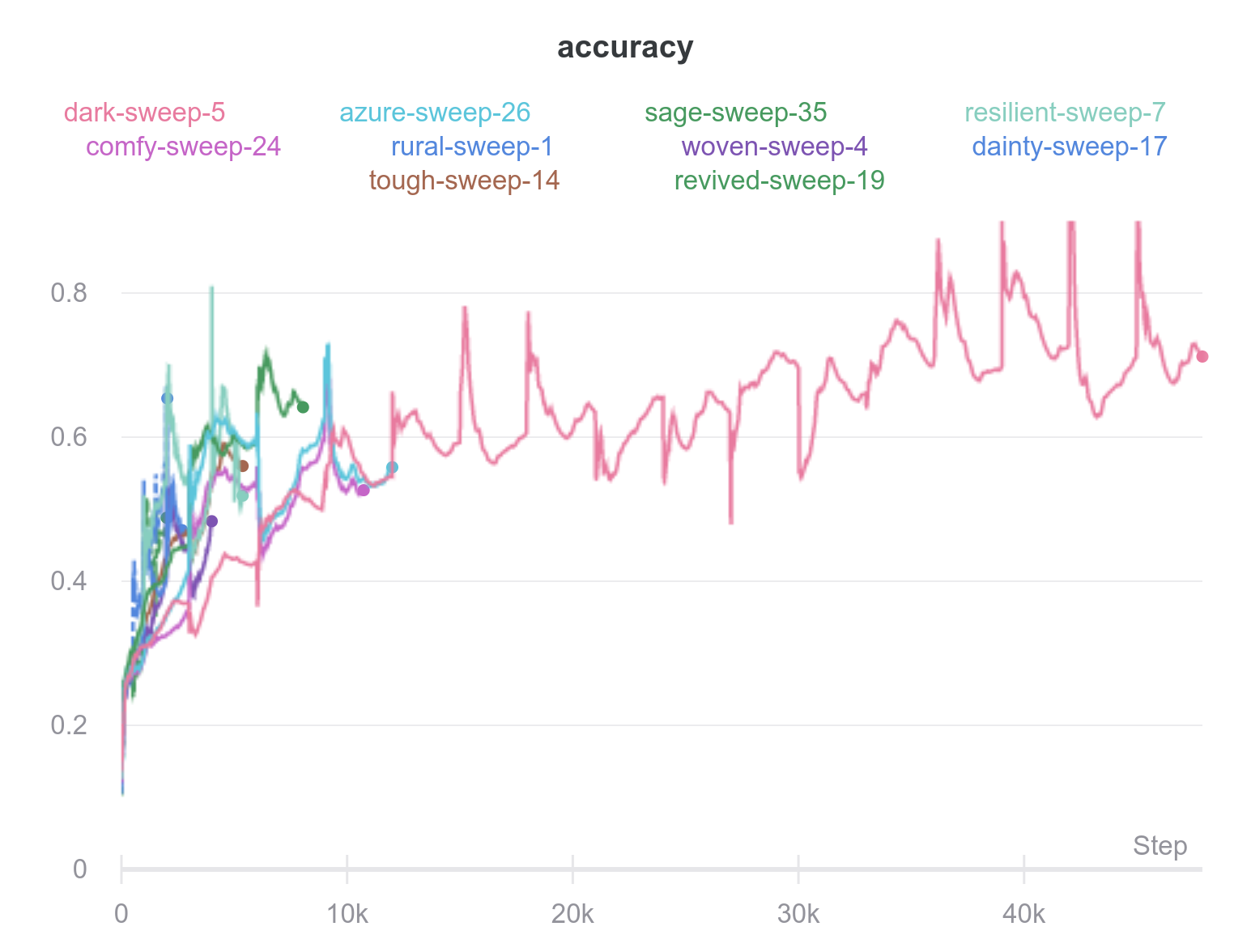}
\includegraphics[width=1.0\linewidth]{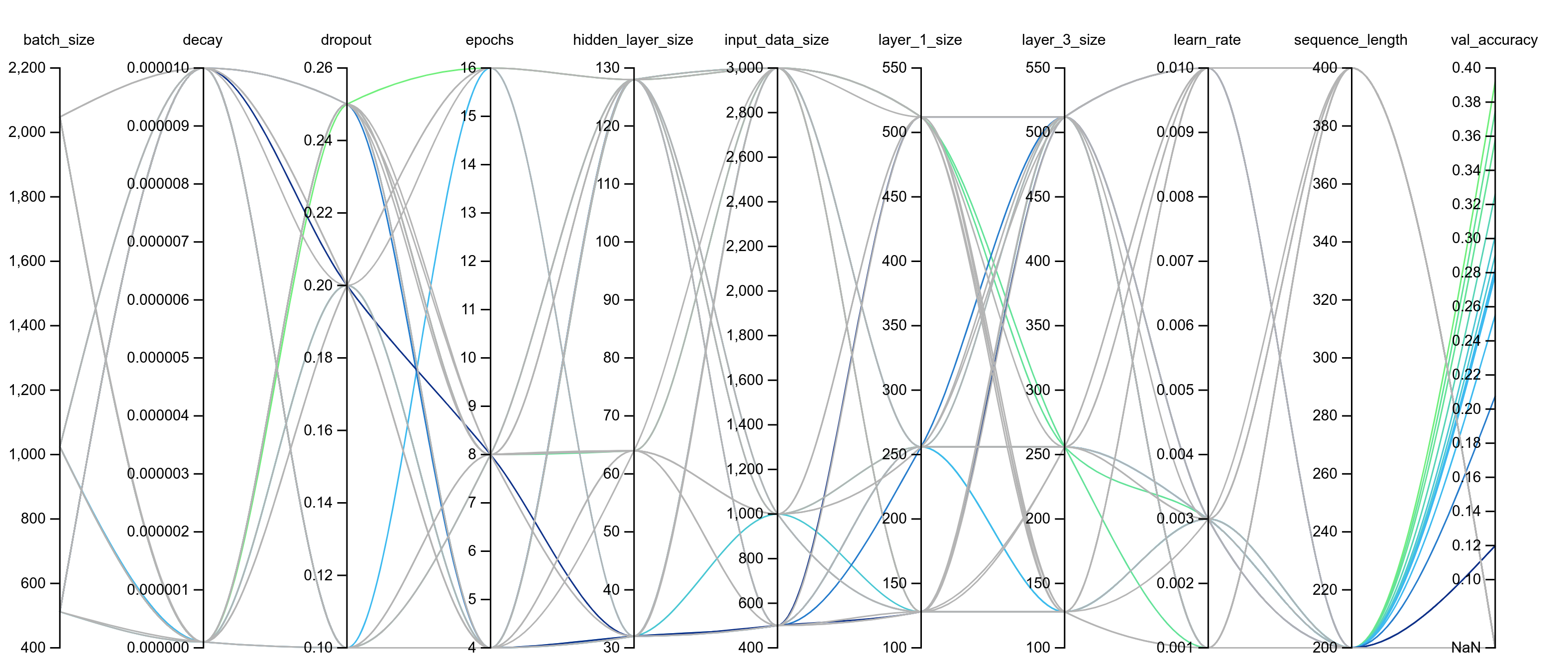}
\end{figure}
\pagebreak
\subsection{Web demo}
A web demo forked from PianoGenie was built using TensorflowJS to load and use the trained model and MagentaJS for the piano player. The demo lets the user play the piano using a MIDI keyboard and use the trained AI model, in the browser!
The interface shows which notes are played by the user and which notes are altered by the AI with different colors. The user is also able to configure the level of performance aid to use through an 'AI confidence' setting.

Please check it out at \url{https://midi-tune.shefa.xyz/}
\begin{figure}[ht]
\centering
\caption{Web demo development}
\includegraphics[width=0.92\linewidth]{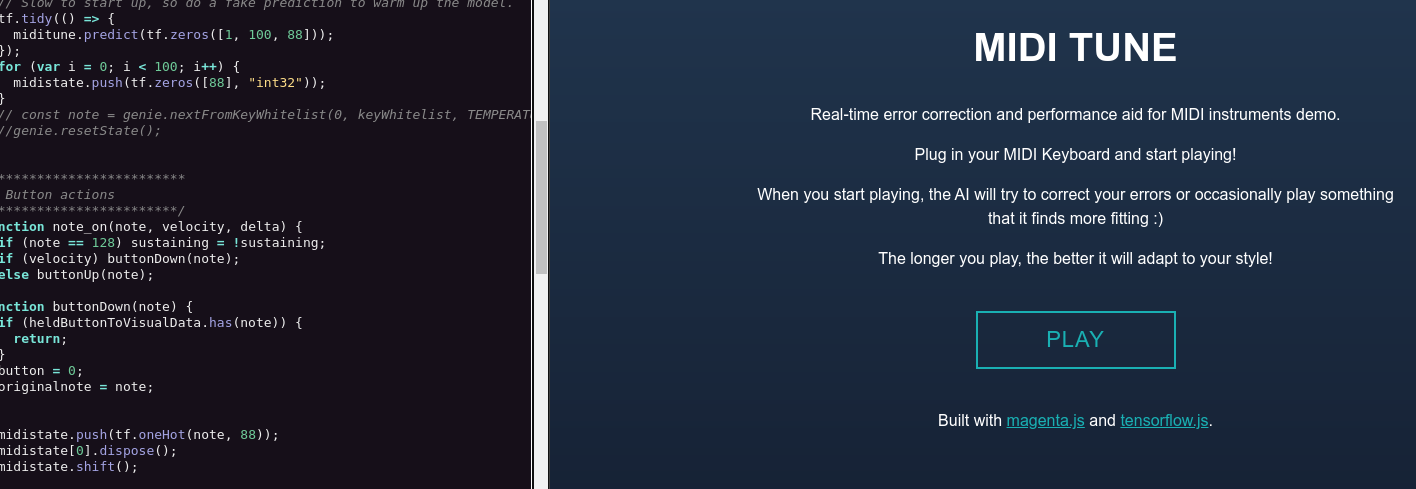}
\end{figure}

\begin{figure}[ht]
\centering
\caption{Orange keys are "played" by the AI, correcting the user's note}
\includegraphics[width=0.92\linewidth]{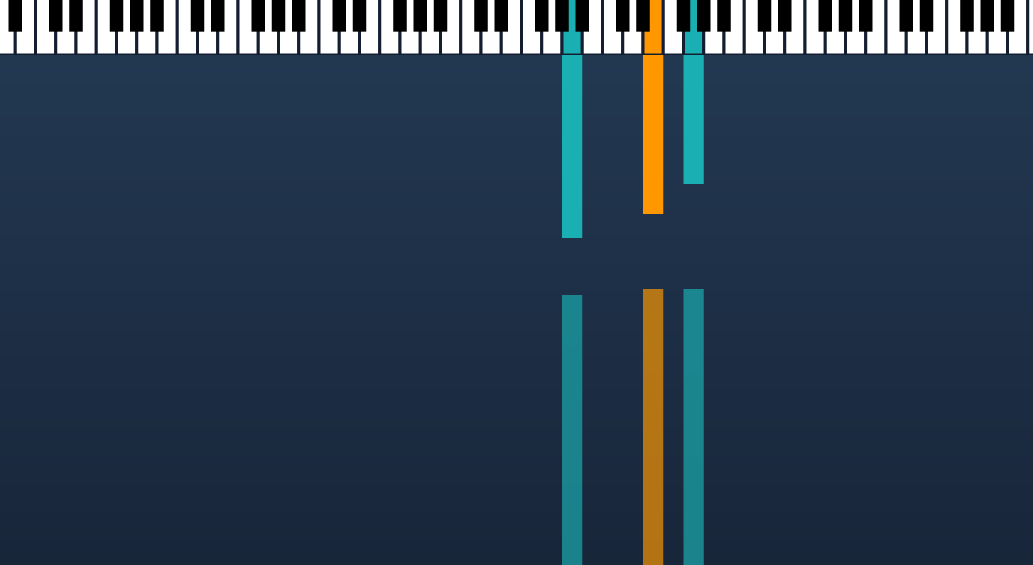}
\end{figure}

\begin{figure}[ht]
\caption{Adjustable performance aid level in the demo}
\includegraphics[width=0.99\linewidth]{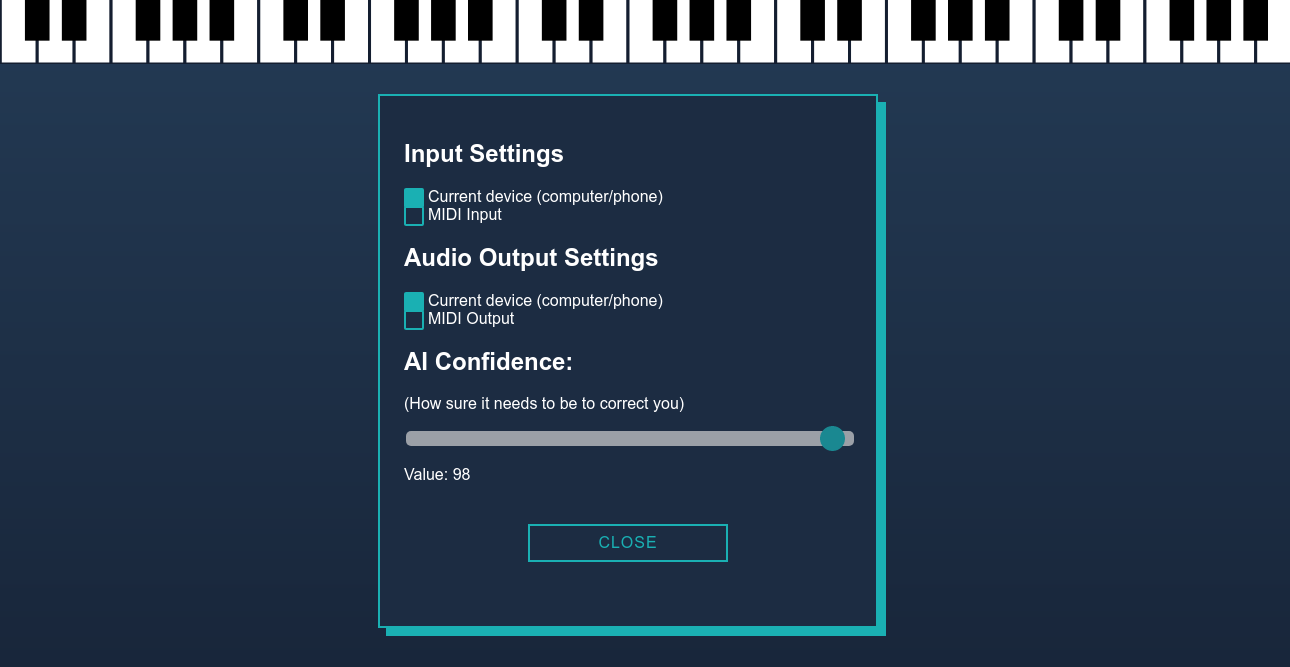}
\end{figure}

\pagebreak

\section{Discussion}
An interesting feature of the developed algorithm is that it operates on the basis of the "intuition" it has formed, and this intuition is not necessarily the same as the player's. Therefore, the musician has to accept that some of the notes that were meant to be heard, will be changed in the course of playing, and will sound differently. There is a trade-off between error correction accuracy and the authenticity of the musical piece performed. This points out to a potential paradigm shift, as the error correction model is better perceived as a performance aid tool, since it also 'corrects' intentionally played notes, which it considers unlikely to occur, with other notes that are more likely to occur or will sound better, according to the AI's taste. 

Note that this characteristic of the AI makes it even more human-like, since if a person heard a piece of music for the first time and had the task to correct errors real-time, she would have corrected not only apparent mistakes, but also notes that were intentionally played by the performer, but she recognized them as a part in the music piece that can be improved. In a way, the developed AI model takes into account the uncertainty that is present in real life!

\pagebreak

\section{Conclusions}

This paper explores some methods for error detection and correction in music performances using MIDI devices. It concludes that real-time auto correction is possible, but identifying only the true mistakes as errors and retrieving the originally intended note is practically impossible due to the artistic nature of music. 

However, the likelihood that a note is played correctly can be estimated, and if not, adjacent notes that are likely to be a better fit within the music context can be played. The final proposed solution is a Recurrent Neural Network, which "plays" along with the performer in real-time by correcting what it considers a mistake or serves the broader purpose of being a performance aid tool. It also allows the user to configure the degree to which the AI can intervene with the original performance. 

Performance aid models of this kind can be applied in various instances, as they allow for more inattentive live playing without depriving it of good quality, and even for collaborative performances between a musician and an "AI musician". The proposed algorithm can be of use to accomplished performers, as well as to novice piano players, to music producers, engineers and composers, as it can not only avoid multiple re-takes, but also help in the generation of new ideas.

The results of the research are available to users in a VST plugin and standalone application, as well as a web demo that can run even on a mobile device. The VST plugin incorporates both context-based error correction and the trained AI model without any noticeable latency. The performance of the web demo could be improved, as there is a slight, but noticeable latency present, especially in mobile.

The research could be further improved by examining other artificial intelligence models, such as the attention based Transformer, when there's enough optimizations to the technique to make it feasible for real-time application.

In conclusion, the machine learning model proposed as a performance aid tool successfully completes the aim of the project and the developed web demo and VST plugin make the findings in this research very accessible to all users.
\pagebreak

\printbibliography

\end{document}

%% file: setup/packages.tex
\usepackage[utf8]{inputenc}
\usepackage[x11names,dvipsnames,svgnames,table]{xcolor}

\usepackage[export]{adjustbox}
\usepackage{afterpage}

\usepackage{graphicx}
\usepackage{placeins}
\usepackage{pdfpages}
\usepackage{algorithm2e}
\usepackage{array}
\usepackage{booktabs}
\usepackage[most]{tcolorbox}
\usepackage{calligra}
\usepackage{caption}
\usepackage{datetime}
\usepackage{dirtytalk}
\usepackage{dsfont}
\usepackage{fancyhdr}
\usepackage{fix-cm}
\usepackage[T1]{fontenc}
\usepackage{textcomp,gensymb} 
\usepackage{graphicx}
\usepackage{lipsum}
\usepackage{listings}
\usepackage{transparent}
\usepackage[everyline=true,framemethod=tikz]{mdframed}
\usepackage{mparhack}
\usepackage{multicol}
\usepackage{multirow}
\usepackage{parskip}
\usepackage{lscape}
\usepackage{pdflscape}
\usepackage{pdfpages}
\usepackage{placeins}
\usepackage[document]{ragged2e}
\usepackage{rotating}
\usepackage{setspace}
\usepackage{subcaption}
\usepackage{threeparttable}
\usepackage[normalem]{ulem}
\usepackage{verbatim}
\usepackage{soul} 

\usepackage[titletoc,title,header]{appendix} 

\usepackage[utf8]{inputenc}
\usepackage[australian]{babel}
\usepackage{csquotes}

\usepackage[a4paper]{geometry}
\usepackage{lastpage} 

\usepackage[hidelinks]{hyperref}
\usepackage{cleveref}

\usepackage{amsmath}
\usepackage{mathtools}

\usepackage{enumitem}
\usepackage{multicol}

\usepackage[style=authoryear-ibid,backend=biber]{biblatex}
\bibliography{bibliography.bib} 

\usepackage[acronym,nonumberlist,toc,section=subsection,numberedsection=nolabel]{glossaries} 

\hypersetup{%
  colorlinks=false,
  linkbordercolor=blue,
  pdfborderstyle={/S/U/W 2}
}